%% file: NY_QV_4_6.tex
\documentclass[12pt]{iopart}
\usepackage[dvips]{epsfig}
\usepackage{amssymb}
\usepackage{amsfonts}
\usepackage{amscd}
\usepackage{pstricks}

 \font\tensym=msbm10
 \font\sevensym=msbm7
 \font\fivesym=msbm5

 \font\tengoth=eufb10
 \font\sevengoth=eufb7
 \font\fivegoth=eufb5

\newfam\symfam
\textfont\symfam=\tensym \scriptfont\symfam=\sevensym
\scriptscriptfont\symfam=\fivesym

\newfam\gothfam
\textfont\gothfam=\tengoth \scriptfont\gothfam=\sevengoth
\scriptscriptfont\gothfam=\fivegoth

\def\hs{\hbox to 3mm{}}
\def\hhs{\hbox to 5cm{}}
\def\ss{\smallskip}
\def\ms{\smallskip}
\def\bs{\bigskip}

\def\JPicScale{0.8}\ifx\JPicScale\undefined\def\JPicScale{1}\fi

\def\A{\mathcal{A}}
\def\Aa{\mathbb{A}}

\def\C{\mathbb{C}}
\def\H{\mathcal{H}}

\def\N{\mathbb{N}}

\def\HWS{{\cal H}_{WS}}
\def\1H{\mathbf{1}_{\HWS}}
\def\L{\mathbb{L}}
\def\V{\mathbb{V}}
\def\X{\mathbb{X}}
\def\Y{\mathbb{Y}}

\def\al{\alpha}
\def\be{\beta}
\def\ep{\varepsilon}

\def\om{\omega}
\def\si{\sigma}

\def\bd{\bar\Delta}
\def\card{\mathrm{card}}
\def\2m#1#2{(#2 #1)}
\def\3m#1#2#3{(#3 #2 #1)}
\def\inf{\uparrow}
\def\deg#1{\mathrm{deg}(#1)}

\def\diag{\mathbf{diag}}
\def\ldiag{\mathbf{ldiag}}
\def\lbell{\mathbf{lbell}}
\def\LDIAG{\mathbf{LDIAG}}

\def\BELL{\mathbf{BELL}}
\def\LBELL{\mathbf{LBELL}}
\def\MQS{\mathbf{MQSym}}
\def\ncp#1#2{#1\langle #2\rangle}

\def\ra{\rightarrow}

\def\ua{\uparrow}

\def\adots{\mathinner{\mkern2mu\raise1pt\hbox{.}
\mkern3mu\raise4pt\hbox{.}\mkern1mu\raise7pt\hbox{.}}}

\def\pointir{\unskip . --- \ignorespaces}

\def\up#1{\raise 1ex\hbox{\footnotesize#1}}

\def\mref#1{{\footnotesize ({\ref{#1}})}}

\newtheorem{expl}{Example}[section]

\newtheorem{theorem}[expl]{Theorem}

\newtheorem{definition}[expl]{D\'efinition}
\newtheorem{proposition}[expl]{Proposition}

\newtheorem{lemma}[expl]{Lemma}
\newtheorem{rem}[expl]{Remark}

\def\Proof{\medskip\noindent {\it Proof --- \ }}
\def\cqfd{\hfill $\Box$ \bigskip}

\begingroup
\def\2#1{\ifnum#1<10 0\fi\the#1}
\xdef\isodayandtime{
{\2\day-\2\month-\the\year\space\2{\count0}:%
\2{\count2}}}
\endgroup

\begin{document}

\title[A Quantum Version of the Hopf Algebra of Feynman-like Diagrams]
{A Three-Parameter Hopf Deformation of the Algebra of Feynman-like
Diagrams}
\author{G H E Duchamp$^{a}$, P Blasiak$^{b}$, A Horzela$^{b}$, K A Penson$^{c}$, A I Solomon$^{c,d}$, \vspace{2mm}}

\address{$^a$ LIPN - UMR 7030\linebreak
CNRS - Universit\'e Paris 13\linebreak
F-93430 Villetaneuse, France\vspace{2mm}}

\address
{$^b$ H. Niewodnicza\'nski Institute of
Nuclear Physics, Polish Academy of Sciences\\
ul. Eliasza-Radzikowskiego 152,  PL 31342 Krak\'ow,
Poland\vspace{2mm}}

\address
{$^c$ Laboratoire de Physique Th\'eorique de la Mati\`{e}re Condens\'{e}e\\
Universit\'e Pierre et Marie Curie, CNRS UMR 7600\\
Tour 24 - 2i\`{e}me \'et., 4 pl. Jussieu, F 75252 Paris Cedex 05,
France\vspace{2mm}}

\address
{$^d$ The Open University, Physics and Astronomy Department\\
Milton Keynes MK7 6AA, United Kingdom\vspace{2mm}}

\eads{\linebreak\mailto{ ghed@lipn-univ.paris13.fr}, \mailto{
a.i.solomon@open.ac.uk} \mailto{ pawel.blasiak@ifj.edu.pl}, \mailto{
andrzej.horzela@ifj.edu.pl}, \mailto{penson@lptl.jussieu.fr}, }

\begin{abstract}
We construct  a three-parameter deformation of the Hopf
algebra $\LDIAG$.  This is the algebra that  appears in an expansion in terms
of Feynman-like diagrams of the {\em product formula} in a simplified
version of Quantum Field Theory. This new algebra is a true Hopf
deformation which reduces to $\LDIAG$ for some parameter values
and to the algebra of Matrix Quasi-Symmetric Functions ($\MQS$) for
others, and thus relates  $\LDIAG$ to other Hopf algebras of
contemporary physics. Moreover, there is an onto linear mapping
preserving products from our algebra to the algebra of Euler-Zagier
sums.
\end{abstract}

\tableofcontents

\newpage

\section{Introduction}

We briefly describe  the passage from the product formula, as described by  by Bender et al. \cite{BBM},  and the related Feynman-like diagrams,  to the description of Hopf algebra structures \cite{GOF4} on the diagrams themselves compatible with their evaluations.\\
First, C. M. Bender, D. C. Brody, and B. K. Meister \cite{BBM} introduced a special field theory which proved to be particularly rich in combinatorial links and by-products.\\
Second, the Feynman-like diagrams produced by this theory label monomials; these monomials  combine in a manner compatible with the monomial multiplication and co-addition\footnote{i.e. the commultiplication obtained by replacing each variable by the sum of two (independent) copies of it.}. This is the Hopf algebra $\mathbf{DIAG}$.\\
Third, the natural noncommutative pull-back of this algebra,
$\LDIAG$, has a basis (the labeled diagrams) which is in one-to-one
correspondence with that of the Matrix Quasi-Symmetric Functions
(the {\em packed matrices} of $\MQS$), but their algebra and co-algebra
structures are completely different. In particular, in this basis,
the multiplication of $\MQS$ implies a sort of {\em shifted shuffle} with
overlappings reminiscent of Hoffmann's shuffle used in the theory of
of polyzeta functions\cite{Ca2}. The superpositions and overlappings
involved there are not present in the (non-deformed) $\LDIAG$ and,
moreover, the coproduct of $\LDIAG$ is co-commutative while that  of
$\MQS$ is not.

\ss
The aim of this paper is to introduce a ``parametric algebra'' which mediates between the two Hopf algebras $\LDIAG$ and $\MQS$. The striking result is that when we  introduce parameters which count  the crossings and overlappings of the shifted shuffle, one notes that the resulting law is associative (graded with unit). We also show how to interpolate with a coproduct which makes, at each stage, our algebra a Hopf algebra. The result is thus a three-parameter Hopf algebra deformation which reduces to $\LDIAG$ at $(0,0,0)$ and to $\MQS$ at $(1,1,1)$. Moreover it appears that, for one set of parameters, the multiplication rule of $\LDIAG$ recovers that of Euler-Zagier sums.\\

{\sc Acknowledgements} : The authors are pleased to acknowledge the hospitality of institutions in Paris, Cracow and New York. Special thanks are due to Catherine Borgen for having created a fertile atmosphere in Exeter (UK) where the first and last parts of this manuscript were prepared. We take advantage of these lines to acknowledge support from the Polish Ministry of Science and Higher Education under Grant N\up{o} 202 107 32/2832. Also, we are grateful to Loic Foissy and Jim Stasheff for their thorough reading.

\section{How and why these Feynman-like Diagrams arise}

The beginning of the story was fully explained in
\cite{GOF2,GOF3,GOF5,GOF6,GOF7,GOF8}, and the Hopf algebra structure
was made precise in \cite{GOF4,GOF10}. In this note we shall  emphasize the latter part of the analysis, where the algebraic structure constructed on the diagrams themselves arise.

\ss Our starting point is the formula ({\em product formula}) of
Bender and al. \cite{BBM}, which can be considered as an expression
of the {\em Hadamard product} for an {\em exponential generating series}. That is,
using

\begin{eqnarray}
F(z)=\sum_{n\geq 0} a_n\frac{z^n}{n!},\ G(z)=\sum_{n\geq 0}
b_n\frac{z^n}{n!},\ \mathcal{H}(F,G):=\sum_{n\geq 0}
a_nb_n\frac{z^n}{n!}
\end{eqnarray}
one can check that

\begin{eqnarray}
\mathcal{H}(F,G)=\left.F\left(z\frac{d}{dx}\right)G(x)\right|_{x=0}.
\end{eqnarray}

When $F(0)$ and $G(0)$ are
not zero one can normalize the functions in this bilinear product so  that $F(0)=G(0)=1$.
We wish  to obtain compact and generic formulas. If we
write the functions as

\begin{eqnarray}
F(z)=\exp\left(\sum_{n=1}^\infty L_n\frac{z^n}{n!}\right),\ \ \ \ \
\ \ G(z)=\exp\left(\sum_{n=1}^\infty V_n\frac{z^n}{n!}\right).
\end{eqnarray}
that is, as free exponentials, then by using   Bell polynomials
in the sets of variables $\L, \V$ (see \cite{GOF4,OPG} for details), we obtain

\begin{eqnarray}
\mathcal{H}(F,G)=\sum_{n\geq 0} \frac{z^n}{n!} \sum_{P_1,P_2\in
UP_n} \L^{Type(P_1)}\V^{Type(P_2)}
\end{eqnarray}
where $UP_n$ is the set of unordered partitions of $[1\cdots n]$. An
unordered partition $P$ of a set $X$ is a subset of $P\subset
\mathfrak{P}(X)-\{\emptyset\}$\footnote{The set of subsets of $X$ is
denoted by $\mathfrak{P}(X)$ (this notation \cite{B_ST} is that of the
former German school).} (that is an unordered collection of blocks,
i. e. non-empty subsets of $X$) such that
\begin{itemize}
    \item the union $\bigcup_{Y\in P}Y=X$ ($P$ is a covering)
    \item $P$ consists of disjoint subsets, i. e.\\
    $Y_1,Y_2\in P\ and\ Y_1\cap Y_2\neq \emptyset \Longrightarrow Y_1=Y_2$.
\end{itemize}

The type of $P\in UP_n$ (denoted above by $Type(P)$) is the multi-index
$(\al_i)_{i\in \N^+}$ such that $\al_k$ is the number of $k$-blocks,
that is the number of members of $P$ with cardinality $k$.

\ss
At this point  the formula entangles and the diagrams of the theory arise.\\
Note particularly that
\begin{itemize}
    \item the monomial $\L^{Type(P_1)}\V^{Type(P_2)}$ needs much less information than that which is contained in the individual partitions $P_1,\ P_2$ (for example, one can relabel the elements without changing the monomial),
    \item two partitions have an incidence matrix {\it from which it is still possible to recover the types of the partitions.}
\end{itemize}

\ss The construction now proceeds as follows.
\begin{enumerate}
    \item Take two unordered partitions of $[1\cdots n]$, say $P_1,P_2$
    \item Write down their incidence matrix $\left(\card(Y\cap Z)\right)_{(Y,Z)\in P_1\times P_2}$
    \item Construct the diagram representing the multiplicities of the incidence matrix : for each block of $P_1$ draw a black spot (resp. for each block of $P_2$ draw a white spot)
    \item Draw lines between the black spot $Y\in P_1$ and the white spot $Z\in P_2$; there are  $\card(Y\cap Z)$ such.
    \item Remove the information of the blocks $Y,Z,\cdots$.
\end{enumerate}

In so doing, one obtains a bipartite graph with $p$ ($=\card(P_1)$)
black spots, $q$ ($=\card(P_2)$) white spots, no isolated vertex and
integer multiplicities. We denote the set of such diagrams by $\diag$.

\bs
\input{unlabelled2_1}

\bs The product formula now reads

\begin{eqnarray}
\mathcal{H}(F,G)=\sum_{n\geq 0} \frac{z^n}{n!} \sum_{d\in diag\atop
|d|=n} mult(d)\L^{\al(d)}\V^{\be(d)}
\end{eqnarray}
where $\al(d)$ (resp. $\be(d)$) is the ``white spots type'' (resp.
the ``black spots type'') i.e. the multi-index $(\al_i)_{i\in \N^+}$
(resp. $(\be_i)_{i\in \N^+}$) such that $\al_i$ (resp. $\be_i$) is
the number of white spots (resp. black spots) of degree $i$ ($i$
lines connected to the spot) and $mult(d)$ is the number of pairs of
unordered partitions of $[1\cdots |d|]$ (here
$|d|=|\al(d)|=|\be(d)|$ is the number of lines of $d$) with
associated diagram $d$.

\ss
Now one may naturally ask \\
{\it Q1) ``Is there a (graphically) natural multiplicative structure
on $\diag$ such that the arrow
\begin{equation}
    d \mapsto \L^{\al(d)}\V^{\be(d)}
\end{equation}
be a morphism ?''}

\ss The answer is ``yes''. The desired product just consists in
concatenating the diagrams (the result, i.e. the diagram obtained
in placing $d_2$ at the right of $d_1$, will be denoted by
$[d_1|d_2]_D$). One must check that this product is compatible with
the equivalence of the permutation of white and black spots among themselves, which is rather straightforward (see \cite{GOF4}). We
have

\begin{proposition} Let $\diag$ be the set of diagrams (including the empty one).\\
i) The law $(d_1,d_2)\mapsto [d_1|d_2]_D$ endows $\diag$ with the structure of a commutative monoid with the empty diagram as neutral element(this diagram will, therefore, be denoted by $1_{\diag}$).\\
ii) The arrow $d \mapsto \L^{\al(d)}\V^{\be(d)}$ is a morphism of
monoids, the codomain of this arrow being the monoid of
(commutative) monomials in the alphabet $\L\cup \V$ i.e.
$$\mathfrak{MON}(\L\cup\V)=\{\L^\al \V^\be\}_{\al,\be\in (\N^+)^{(\N)}}=
\bigcup_{n,m\geq 1}\big\{L_1^{\al_1}L_2^{\al_2}\cdots
L_n^{\al_n}V_1^{\be_1}V_2^{\be_2}\cdots
V_m^{\be_m}\big\}_{\al_i,\be_j\in \N}.$$ iii) The monoid
$(\diag,[-|-]_D,1_{\diag})$ is a free commutative monoid. Its
letters are the connected (non-empty) diagrams.
\end{proposition}

\begin{rem} The reader who is not familiar with the algebraic structure of $\mathfrak{MON}(\X)$ can find rigorous definitions in paragraph \mref{fcmon} where this structure is needed for the proofs relating to deformations.
\end{rem}

\section{Non-commutative lifting (classical case)}

The ``classical'' construction of the Hopf algebra $\LDIAG$ was
given in \cite{GOF4}. We give  the proofs below, using a {\em coding}
through ``lists of monomials'' needed for the deformed (quantum)
case. The entries of a list can be considered as ``coordinate
functions'' for the diagrams (see introduction of section
\mref{classical}).

\subsection{Free monoids}\label{fcmon}

We  recall here the construction of the free and free-commutative monoids generated by a given set of variables (i.e. an alphabet) \cite{BR}.\\
Let $\X$, be a set. We denote by $\X^*$ the set of lists of elements
of $\X$, including the empty one. In many works, and in the sequel,
the list $[x_1,x_2,\cdots ,x_n]$ will be considered as a word
$x_1x_2\cdots x_n$ so that the concatanation of two lists
$[x_1,x_2,\cdots ,x_n],\ [y_1,y_2,\cdots ,y_m]$ is just the word
$x_1x_2\cdots x_ny_1y_2\cdots y_m$. For this (associative) law, the
empty list $[\ ]$ is the neutral element and will therefore be denoted by $1_{\X^*}$

\ss Similarly, we denote by $\N^{(\X)}$ \cite{B_Alg_III} the set of
multisubsets of $\X$ (i.e. the set of - multiplicity - mappings with
finite support $\X\mapsto \N$). Every element $\al$ of $\N^{(X)}$
can be written multiplicatively, following the classical multi-index
notation
\begin{equation}
    \X^{\al}=\prod_{x\in \X} x^{\al(x)}
\end{equation}
and the set $\mathfrak{MON}(X)=\{\X^{\al}\}_{\al\in \N^{(X)}}$ is
exactly the set of (commutative) monomials with variables in $\X$.
It is a monoid, indeed a (multiplicative) copy of $\N^{(X)}$ as
$\X^\al\X^\be=\X^{\al+\be}$. The subset of its non-unit elements is
a semigroup which will be denoted by $\mathfrak{MON}^+(X)$
($=\mathfrak{MON}(X)-\{\X^0\}$).

\subsection{Labeling the nodes}

There are (at least) two good reasons to look for non-commutative
structures which may serve as a noncommutative pullback for $\diag$.
\begin{itemize}
    \item Rows and Columns of matrices are usually (linearly) ordered and we have seen that a diagram is not represented by a matrix but by a class of matrices
    \item The complexity of $\diag$ and its algebra is not sufficient to relate it to other (non-commutative or non-cocommutative) algebras relevant to contemporary physics
\end{itemize}

The solution (of the non-deformed problem \cite{GOF4}) is simple and
consists in labeling the nodes from left to right and from ``$1$''
to the desired number as follows.

\bs
\input{labelled2_1}

\bs The set of these graphs (i.e. bipartite graphs on some product
$[1..p]\times [1..q]$ with no isolated vertex) will be denoted by
$\ldiag$. The composition law is, as previously, concatenation
in the obvious sense. Explicitly, if $d_i,\ i=1,2$ are two diagrams
of dimension $[1..p_i]\times [1..q_i]$, one relabels the black
(resp. white) spots of $d_2$ from $p_1+1$ to $p_1+p_2$ (resp. from
$q_1+1$ to $q_1+q_2$) the result will be noted $[d_1|d_2]_L$. One
has

\begin{proposition} Let $\ldiag$ be the set of labeled diagrams (including the empty one).\\
i) The law $(d_1,d_2)\mapsto [d_1|d_2]_L$ endows $\ldiag$ with the structure of a noncommutative monoid with the empty diagram ($p=q=0$) as neutral element(which will, therefore, be denoted by $1_{\ldiag}$).\\
ii) The arrow from $\ldiag$ to $\diag$, which implies ``forgetting the labels of the vertices'' is a morphism of monoids.\\
iii) The monoid $(\ldiag,[-|-]_L,1_{\ldiag})$ is a free
(noncommutative) monoid. Its letters are the irreducible diagrams
(denoted from now on by $irr(\ldiag)$).
\end{proposition}

\begin{rem}
i) In a general monoid $(M,\star,1_M)$, the irreducible elements are the elements $x\neq 1_M$ such that $x=y\star z\Longrightarrow 1_M\in \{y,z\}$.\\
ii) It can happen that an irreducible of $\ldiag$ has an image in
$\diag$ which splits, as shown by  the simple example of the {\em cross}
defined by the incidence matrix ${\pmatrix{0 & 1\cr 1 & 0}}$.
\end{rem}

\subsection{Coding $\ldiag$ with ``lists of monomials''}

One can code every labelled diagram by a ``list of (commutative)
monomials'' in the following way.
\begin{itemize}
    \item Let $\X=\{x_i\}_{i\geq 1}$ be an infinite set of indeterminates and $d\in \ldiag_{p\times q}$ a diagram ($\ldiag_{p\times q}$ is the set of diagrams with $p$ black spots and $q$ white spots).
    \item Associate with $d$ the multiplicity function $[1..p]\times [1..q]\ra \N$ such that $d(i,j)$ is the number of lines from the black spot $i$ to the white spot $j$.
    \item The code associated with $d$ is $\varphi_{lm}(d)=[m_1,m_2,\cdots ,m_p]$ such that $m_i=\prod_{j=1}^q x_j^{m(i,j)}$
\end{itemize}
\noindent

\bs
\input{coding1}

\bs As a data structure, the lists of monomials are elements of
$(\mathfrak{MON}^+(X))^*$, the free monoid whose letters are
$\mathfrak{MON}^+(X)=\mathfrak{MON}(X)-\{\X^0\}$, the semigroup of non-unit
monomials over $\X$.

\ss It is not difficult to see that, through this coding,
concatenation is reflected in the following formula
\begin{equation}\label{sconc}
    \varphi_{lm}([d_1|d_2]_L)=\varphi_{lm}(d_1)\ast T_{max(IndAlph(\varphi_{lm}(l_1)))}(\varphi_{lm}(d_2))
\end{equation}
where $T_{p}$ is the translation operator which changes the variables according to\\
$T_p(x_i)=x_{i+p}$ (which corresponds to the relabelling of the white spots) and $p_1$ is the number of black spots of $d_1$.\\
For example, one has
\begin{equation}
    T_2([x_2^2x_3,x_1x_2x_3^3,x_3x_4^2])=[x_4^2x_5,x_3x_4x_5^3,x_5x_6^2]\ ;\ T_6([x_1,x_2^2])=[x_7,x_8^2]
\end{equation}

\section{The Hopf algebra $\LDIAG$ (non-deformed case)}\label{classical}

In \cite{GOF4}, we defined a Hopf algebra structure
on the space of diagrams $\LDIAG$. The aim of this section is to
give complete proofs and details for this construction through the
use of the special space of coordinates constructed above (the
complete vector  of coordinates of a diagram being its code).

\subsection{The monoid $(\mathfrak{MON}^+(X))^*$ and the submonoid of codes of diagrams}

Formula \mref{sconc} can be written using  lists as

\begin{equation}\label{lists_sconc}
l_1\bar\ast l_2=l_1\ast T_{max(IndAlph(l_1))}(l_2)
\end{equation}

which defines a monoid structure on $(\mathfrak{MON}^+(X))^*$ (the set of lists of non-unit monomials) with the empty list as neutral (i.e. $[\ ]$ which will, therefore, be denoted by ``$1_{(\mathfrak{MON}^+(X))^*}$'' or simply ``$1$'' when the context is clear).\\
We will return to this construction (called shifting \cite{FPSAC07}) later.\\
The alphabet of a list is the set of variables occurring in the list.
Formally
\begin{equation}
    Alph([m_1,m_2,\cdots m_n])=\bigcup_{1\leq i\leq k} Alph(m_i)
\end{equation}
where, classically, for a monomial $m=\X^{\al}$, $Alph(m)=\{x_i\}_{\al(i)\not= 0}$.\\
Now, we can define the ``compacting operator'' on
$\ncp{k}{\mathfrak{MON}^+(X)}$ by its action on the lists. This operator
actually removes the holes in the alphabet of a list by pushing to
the left the indices which are at the right of a hole. For example
(we denote by $cpt$ the operator)

\begin{equation}
cpt([x_2^2x_{10},x_3x_4x_8^3,x_3x_4^2])=[x_1^2x_5,x_2x_3x_4^3,x_2x_3^2].
\end{equation}

The alphabet of the list on the LHS is
$Alph(l)=Alph([x_2^2x_{10},x_3x_4x_8^3,x_3x_4^2])=\{x_2,x_3,x_4,x_8,x_{10}\}$,
its indices are $IndAlph(l)=\{2,3,4,8,10\}$ and the re-indexing
function is the unique strictly increasing mapping from
$\{2,3,4,8,10\}$ to $[[5]]$. Here the compacting operator is just
the substitution
$$
x_1\leftarrow x_2;\ x_2\leftarrow x_3;\ x_3\leftarrow x_4;\
x_4\leftarrow x_8;\ x_5\leftarrow x_{10}
$$

The formal definitions are the following
\begin{itemize}
    \item $IndAlph(l)=\{i\ |\ x_i\in Alph(l)\}$
    \item $l$ being given, let $\phi_l$ be the unique increasing mapping from $IndAlph(l)$ to $[[card(IndAlph(l))]]$ (in fact, $card(IndAlph(l))=card(Alph(l))$)
    \item let $s_l$ be the substitution $x_i\leftarrow x_{\phi_l(i)}$ in the monomials.
    \item Then, if $l=[m_1,m_2,\cdots m_n]$, $cpt(l)=[s_l(m_1),s_l(m_2),\cdots s_l(m_n)]$.
\end{itemize}

\begin{definition}
The compacting operator $cpt\ :\ \ncp{k}{\mathfrak{MON}^+(X)}\mapsto
\ncp{k}{\mathfrak{MON}^+(X)}$ is the extension by linearity of the
mapping $cpt$ defined above.
\end{definition}

It can be checked easily that, for $l\in (\mathfrak{MON}^+(X))^*$, the
following are equivalent
\begin{enumerate}
\item $cpt(l)=l$
\item $IndAlph(l)=[[card(IndAlph(l))]]$
\item there is no hole in $Alph(l)$; that is, there exists no $i\geq 1$ s.t. $x_i\notin Alph(l)$ and $x_{i+1}\in Alph(l))$
\item $l$ is the code of some (then unique) diagram $d$.
\end{enumerate}

\ss It follows from the preceding properties that $cpt$ is a
projector with range the subspace $\mathcal{C}_{ldiag}$ of
$\ncp{k}{\mathfrak{MON}^+(\X)}$ generated by the codes of the diagrams.
Formula \mref{sconc} proves that $\mathcal{C}_{ldiag}$ is closed
under the shifted concatenation defined by \mref{lists_sconc}. More
precisely

\begin{proposition}
The algebra $\mathcal{C}_{ldiag}$ is a free algebra on the set of
the codes of irreducible diagrams.
\end{proposition}

These codes are also the non-empty lists $l$ which are compact (i.e.
$cpt(l)=l$) and cannot be factorized into a product of two non-empty
lists i.e. $l=l_1*l_2;\ l_i\not= [\ ]$ (one can check easily that,
if $l_1*l_2$ is compact, so are $l_1$ and $l_2$).

\subsection{The Hopf algebras $\mathcal{C}_{ldiag}$ and $\LDIAG$}

The algebra $\LDIAG$ is endowed with the structure of a bi-algebra by
the comultiplication

\begin{equation}
    \Delta_{BS}(d)=\sum_{I+J=[1..p]} d[I]\otimes d[J]
\end{equation}
where $p$ is the number of black spots and $d[I]$ is the ``restriction'' of $d$ to the black spots selected by the $I\subset [1..p]$.\\
On the other hand, we have a standard Hopf algebra structure on
the free algebra, expressed in terms of concatenation and subwords
\cite{Ha,Re}. Let $\Aa$ be an alphabet (a set of letters) and $w\in
\Aa^*$ a word, if we write $w$ a a sequence of letters
$w=a_1a_2\cdots a_n;\ a_i\in \Aa$, the length $|w|$ of $w$ is $n$
and if $I=\{i_1,i_2,\cdots i_k\}\subset [1..n]$, the subword $w[I]$
is $a_{i_1}a_{i_2}\cdots a_{i_k}$ (this notation is slightly
different from that of \cite{Re} where it is $w|_I$). Then, the free
algebra $\ncp{k}{\Aa}$ is a Hopf algebra with comultiplication
\cite{Re,Ha}.

\begin{equation}\label{liehopf}
    \Delta_{LieHopf}(w)=\sum_{I+J=[1..n]}w[I]\otimes w[J].
\end{equation}

One has the following relation between restrictions of diagrams and
subwords
\begin{equation}
    \varphi_{lm}(d[I])=cpt(\varphi_{lm}(d)[I])
\end{equation}
this suggests that the coproduct
\begin{equation}
    \Delta_{list}(l)=\sum_{I+J=[1..n]} cpt(l[I])\otimes cpt(l[J])
\end{equation}
could be a Hopf algebra comultiplication for the shifted algebra
$(\ncp{k}{\mathfrak{MON}^+(\X)},\bar\ast,[\ ])$. Unfortunately, this
fails due to the lack of counit (i and ii of the following Theorem),
but the ``ground subalgebra''  $\mathcal{C}_{ldiag}$ is a genuine
Hopf algebra (which is exactly what we do need here).

\begin{theorem}
Let $\A=(\ncp{k}{\mathfrak{MON}^+(\X)},\bar\ast,[\ ])$ be the algebra of lists of (non-unit) monomials endowed with the shifted concatenation of formula \mref{lists_sconc}. Then\\
i) $\A$ is a free algebra.\\
ii) The coproduct $\Delta_{list}$ (recalled below) is co-associative and a morphism of algebras $\A\mapsto \A\otimes \A$ (i.e. $\A$ is a bi-algebra without counit).\\

\begin{equation}
    \Delta_{list}(l)=\sum_{I+J=[1..n]} cpt(l[I])\otimes cpt(l[J])
\end{equation}

iii) The algebra $\mathcal{C}_{ldiag}$ is a sub-algebra and
coalgebra of $\A$ which is a Hopf algebra for the following co-unit
and antipode.

\begin{itemize}
    \item {\sc Counit}
\begin{equation}
    \ep(l)=\delta_{l,[\ ]} \mathrm{\hspace{2cm }(Kronecker\ delta)}
\end{equation}
\item {\sc Antipode}
\begin{equation}
    S(l)=
\sum_{r\geq 0}\sum_{I_1+I_2+\dots I_r=[1..p]\atop I_j\not=
\emptyset} (-1)^r cpt(l[I_1])cpt(l[I_2])\cdots cpt(l[I_r])
\end{equation}
\end{itemize}
\end{theorem}

\Proof i) Throughout the proof, we will denote by $\ast$ the
concatenation between lists and $\bar\ast$ the shifted concatenation
defined by the formula \mref{lists_sconc}. We first remark that, if
$l=l_1\bar\ast l_2$, then $max(IndAlph(l_1))<min(IndAlph(l_2))$.
This leads us to define, for a (non-shifted) factorization
$l=l_1\ast l_2=l[1..t]\ast l[t+1..p]$ ($p=|l|$), a gauge of the
degree of overlapping of the intervals (of integers)
$[1..max(IndAlph(l1))]$ and $[min(IndAlph(l2))..\infty[$, thus the
function

\begin{eqnarray}
\om_l(t)=card\bigg([1..max(IndAlph(l[1..t])]\cap
[min(IndAlph(l[t+1..p])..\infty[\bigg)=\cr
\bigg(max(IndAlph(l[1..t]))-min(IndAlph(l[t+1..p]))+1\bigg)^{+} .
\end{eqnarray}
(We recall that, for a real number $x$, $x^+$ is its positive part
$x^+=max(x,0)=\frac{1}{2}(|x|+x)$ \cite{B_Alg_VI}). It can be
easily checked that the points $t$ where $\om_l(t)=0$ determine the
(unique) factorisation of $l$ in irreducibles. It follows that
the monoid $((\mathfrak{MON}^+(\X))^*,\bar\ast,[\ ])$ is free and so is
its algebra $(\ncp{k}{\mathfrak{MON}^+(\X)},\bar\ast,[\ ])$.

\ss ii) If we denote $\Delta : \A\mapsto \A\otimes \A$ the standard
coproduct given, for a list $l$ of length $p$, by formula
\mref{liehopf}, one can remark that

\begin{enumerate}
    \item $cpt(l_1)\bar\ast cpt(l_2)=cpt(l_1\bar\ast l_2)$
    \item $\Delta_{list}=(cpt\otimes cpt)\circ \Delta$
    \item $\Delta_{list}\circ cpt=\Delta_{list}$
    \item $(\forall n\in \N) (cpt(T_n(l))=cpt(l))$
    \item $(\forall n\in \N) (\Delta\circ T_n=(T_n\otimes T_n)\circ \Delta)$.
\end{enumerate}

{\sc Coassociativity of $\Delta_{list}$}\pointir

\bs One has

\begin{eqnarray}
(\Delta_{list}\otimes Id)\circ\Delta_{list}=(\Delta_{list}\otimes
Id)\circ (cpt\otimes cpt)\circ \Delta=`\cr ((\Delta_{list}\circ
cpt)\otimes cpt)\circ \Delta= (\Delta_{list}\otimes cpt)\circ
\Delta=\cr (((cpt\otimes cpt)\circ \Delta)\otimes cpt)\circ
\Delta=\cr (cpt\otimes cpt\otimes cpt)\circ (\Delta\otimes Id)\circ
\Delta= (cpt\otimes cpt\otimes cpt)\circ (Id\otimes \Delta)\circ
\Delta\cr (cpt\otimes ((cpt\otimes cpt)\circ \Delta))\circ
\Delta=(cpt\otimes \Delta_{list})\circ \Delta=\cr (cpt\otimes
(\Delta_{list}\circ cpt))\circ \Delta=\cr (Id\otimes
\Delta_{list})\circ (cpt\otimes cpt)\circ \Delta= (Id\otimes
\Delta_{list})\circ \Delta_{list}
\end{eqnarray}
{\sc $\Delta_{list}$ is a morphism}\pointir

\bs For two lists $u,v\in$, let us compute $\Delta_{list}(u\bar\ast
v)$. With $p=max(IndAlph(u))$, one has

\begin{eqnarray}\label{morph_init}
\Delta_{list}(u\bar\ast v)=(cpt\otimes cpt)\circ \Delta(l_1\ast
T_p(v))=\cr (cpt\otimes cpt)(\Delta(u)\ast^{\otimes 2}
\Delta(T_p(v)))= \cr (cpt\otimes cpt)(\Delta(u)\ast^{\otimes 2}
(T_p\otimes T_p)\Delta (v)=  \cr (cpt\otimes cpt) (\sum_{(1)(2)}
u_{(1)}\otimes u_{(2)})\ast^{\otimes 2} (T_p\otimes T_p)
(\sum_{(3)(4)} v_{(3)}\otimes v_{(4)})=\cr (cpt\otimes cpt)
(\sum_{(1)(2)(3)(4)} u_{(1)}\ast T_{p_1}(T_{p-p_1}(v_{(3)}))\otimes
u_{(2)}\ast T_{p_2}(T_{p-p_2}(v_{(4)})))
\end{eqnarray}
with, for each term in the sum
$$
p_1=max(IndAlph(u_{(1)}))\leq p\ ;\ p_2=max(IndAlph(u_{(2)}))\leq p
$$
so, the quantity in \mref{morph_init} is
\begin{eqnarray}
(cpt\otimes cpt) (\sum_{(1)(2)(3)(4)} u_{(1)}\bar\ast
(T_{p-p_1}(v_{(3)}))\otimes u_{(2)}\bar\ast
(T_{p-p_2}(v_{(4)})))=\cr \sum_{(1)(2)(3)(4)} cpt(u_{(1)}\bar\ast
(T_{p-p_1}(v_{(3)})))\otimes cpt(u_{(2)}\bar\ast
(T_{p-p_2}(v_{(4)})))=\cr \sum_{(1)(2)(3)(4)}
\bigg(cpt(u_{(1)})\bar\ast cpt(T_{p-p_1}(v_{(3)}))\bigg)\otimes
\bigg(cpt(u_{(2)})\bar\ast cpt(T_{p-p_2}(v_{(4)}))\bigg)=\cr
\sum_{(1)(2)(3)(4)} \bigg(cpt(u_{(1)})\bar\ast
cpt(v_{(3)})\bigg)\otimes \bigg(cpt(u_{(2)})\bar\ast
cpt(v_{(4)})\bigg)=\cr \bigg(\sum_{(1)(2)} cpt(u_{(1)})\otimes
cpt(u_{(2)})\bigg)\bar\ast^{\otimes 2} \bigg(\sum_{(3)(4)}
cpt(v_{(3)})\otimes cpt(v_{(4)})\bigg)=\cr
\Delta_{list}(u)\bar\ast^{\otimes 2} \Delta_{list}(v)
\end{eqnarray}

\bs iii) As $\mathcal{C}_{ldiag}$ is generated by the image of $cpt$
it is clear that this space is a sub-coalgebra of $\A$. Moreover,
$cpt$ is a (multiplicative) morphism $\A\mapsto \A$ and thus its
image $\mathcal{C}_{ldiag}$ is a subalgebra of $\A$. We now supply the missing ingredients to complete the proof of the Hopf algebra structure.

\bs {\sc $\ep$ is a counit}\pointir

\ss Let $l=cpt(l)$ be a compact list. We remark that, for any list
$u$, one has $cpt(u)=[\ ]\Longleftrightarrow u=[\ ]$. Then, with
$\mu_l\ :\ k\otimes \A\mapsto \A$ the scaling operator

\begin{eqnarray}
\mu_l(\ep\otimes    Id)\Delta_{list}(l)=\sum_{I+J=[1..n]}
\ep(cpt(l[I]))cpt(l[J])=\cr \hspace{-2cm}\sum_{I+J=[1..n]\atop
I=\emptyset} \ep(cpt(l[I]))cpt(l[J]) + \sum_{I+J=[1..n]\atop
I\not=\emptyset} \ep(cpt(l[I]))cpt(l[J])=cpt(l)+0=l
\end{eqnarray}
the proof of the fact that $\ep$ is a left counit is similar.

\bs {\sc $S$ is the antipode}\pointir

\ss One has $\mathcal{C}_{ldiag}=k.1\oplus ker(\ep)$, let us denote
$Id^+$ the projection
$\mathcal{C}_{ldiag}\mapsto ker(\ep)$ according to this decomposition.\\
Then, for every list $l$,
$$
\sum_{r\geq 0}\sum_{I_1+I_2+\dots I_r=[1..p]\atop I_j\not=
\emptyset} (-1)^r cpt(l[I_1])cpt(l[I_2])\cdots cpt(l[I_r])
$$
is well defined as the first sum is locally finite. Thus, the
operator
$$
\sum_{r\geq 0}\sum_{I_1+I_2+\dots I_r=[1..p]\atop I_j\not=
\emptyset} (-1)^r \underbrace{(Id^+*Id^+*\cdots *Id^+)}_{r\ times}
$$
is well defined and is the convolutional inverse of $Id$.

\subsection{Subalgebras of $\LDIAG$}

\subsubsection{Graphic primitive elements}\label{GPE}

\bs
The problem of Graphic Primitive Elements (GPE) is the following.\\
Let $\H$ be a Hopf algebra with (linear) basis $G$, a set of graphs.
The GPE are the primitive elements $\Gamma\in G$ which are primitive
i.e.
\begin{equation}
\Gamma  \textrm{ is a GPE} \Longleftrightarrow \Gamma\in G \textrm{
and } \Delta(\Gamma)=\Gamma\otimes 1+1\otimes \Gamma.
\end{equation}
It is not difficult to check that, in any case, the subalgebra $\H^{\textrm{\tiny GPE}}$ generated by these elements is also a sub-coalgebra.\\
We make an extra hypothesis (which is often fulfilled)
\begin{equation}\label{extrahyp}
    1_\H\in G \textrm{ and } (\Gamma\in G-\{1_\H\}\Longrightarrow \ep(\Gamma)=0).
\end{equation}
Then (if \mref{extrahyp} is fulfilled) $\H^{\textrm{\tiny GPE}}$ is
a sub-Hopf algebra as the antipode of the product
$\Gamma_1\Gamma_2\cdots \Gamma_p$ of (GPE) is
\begin{equation}
    S(\Gamma_1\Gamma_2\cdots \Gamma_p)=(-1)^p\ \Gamma_p\Gamma_{p-1}\cdots \Gamma_1.
\end{equation}
The following proposition helps to determine $\LDIAG^{\textrm{\tiny
GPE}}$.

\begin{proposition} In $\LDIAG$ (with basis $G=\ldiag$), the following are equivalent\\
i) $d$ is a GPE\\
ii) $d$ has only one black spot.
\end{proposition}

Then, the Hopf algebra $\LDIAG^{\textrm{\tiny GPE}}$ is generated by
the product of ``one-black-spot'' diagrams.

\bs
\input{GPE2}

\bs

\subsubsection{Level subalgebras}\hspace{1mm}

\bs
One can also impose  limitations on the incoming degrees of the white spots in a way compatible with the coproduct. In this case, one defines an infinity of Hopf-subalgebras of $\LDIAG$ which we will call ``level subalgebras''.\\
More precisely, given an integer $l>0$, one can ask for spaces
generated by the diagrams $d$ for which every white spot has an
incoming degree $\leq l$. This amounts to say that the ``white spot
type'' of every diagram $d$ is of the form
$$
\al(d)=(\al_1,\al_2,\cdots \al_k,0,0\cdots 0,\cdots)\ ;\ (\textrm{
all the } \al_i\leq l\textrm{ for }i\leq k\textrm{ and
}\al_i=0\textrm{ for }i>k)
$$
We denote by $\LDIAG^{\leq l}$ the subspace generated by these
diagrams. One has a chain of Hopf algebras

\begin{equation}
\hspace{-1.5cm} \LDIAG^{\leq 1}\subset \LDIAG^{\leq 2}\subset \cdots
\LDIAG^{\leq l}\subset \LDIAG^{\leq l+1}\subset\cdots  \subset\LDIAG
\end{equation}

In the next paragraph, we will specially be interested in
$$
\LBELL=\LDIAG^{\leq 1}\cap \LDIAG^{\textrm{\tiny GPE}}.
$$

\bs
\subsubsection{$\BELL$ and $\LBELL$}\hspace{1mm}

\bs
The algebras $\BELL$ and $\LBELL$ were defined in \cite{GOF9}.\\

The algebra $\LBELL$ is the intersection $\LDIAG^{\leq 1}\cap
\LDIAG^{\textrm{\tiny GPE}}$ and since they are subspaces
generated by subsets of $\ldiag$, $\LBELL$ is generated by diagrams
that
\begin{itemize}
    \item are concatenations of one-black-spot-diagrams
    \item such that the incoming degree of every white spot is one.
\end{itemize}
Let $d_k$ be the diagram with code $[x_1,x_2,\cdots x_k]$. $\LBELL$
is generated by concatenations of these diagrams. Indeed, the
diagrams $d_k$ are a subalphabet of the free monoid $\ldiag$ so that
they generate a free submonoid which we will denote here $\lbell$.

\bs
\input{lbell}

\bs The algebras $\LDIAG$ and $\LBELL$ are both enveloping
algebras. They are generated by their primitive elements which are in
general linear combinations of diagrams and not pure diagrams. For an
analysis of ``graphic primitive elements'' see section \mref{GPE}.

\section{The algebra $\LDIAG(q_c,q_s,q_t)$ (deformed case)}

\subsection{Counting crossings ($q_c$) and superpositions ($q_s$)}

The preceding coding is particularly well adapted to the the
deformation we want to construct here. The philosophy of the
deformed product is expressed by the descriptive formula\footnote{Exact definition of the coefficient $q_c^{nc\times weight}q_s^{weight\times weight}$ is the result of crossing and shifting processes which will be detailed in paragraph \mref{modified}.}.
\begin{equation}
    [d_1|d_2]_{L(q_c,q_s)}=\sum_{cs(?)\ all\ crossing\ and\atop superpositions\ of\ black\ spots}
    q_c^{nc\times weight}q_s^{weight\times weight} cs([d_1|d_2]_{L})
\end{equation}
where
\begin{itemize}
    \item $q_c,q_s\in \C$ or $q_c,q_s$ formal. These and other cases may be unified by considering the set of coefficients as belonging to a ring $K$.
    \item the exponent of $q_c^{nc\times weight}$ is the number of crossings of ``what crosses'' times its weight
    \item the exponent of $q_s^{weight\times weight}$ is the product of the weights of ``what is overlapped''
    \item $cs( )$ are the diagrams obtained from $[d_1|d_2]_L$ by the process of crossing and superposing the black spots of $d_2$ on to those of $d_1$, the order and distinguishability of the black spots of $d_1$ (i.e. $d_2$) being preserved.
\end{itemize}

What is striking is that this law is associative. This result will
be established after the following paragraph.

\bs
\input{cross1}

\bs
\input{detail2}

\subsection{Modified laws}\label{modified}

\subsection*{$\bullet$ Twisting}

\begin{proposition}\label{2params} Let $A=(A_n)_{n\in \N}$ a graded semigroup and $A^*$ the set of lists
(denoted by $[a_1, a_2, \cdots a_k]$) with letters in $A$.\\
For convenience we define the operator $\ast$ (left append) $A\times
A^*\mapsto A^*$ by
\begin{equation}
a\ast [b_1, b_2, \cdots b_n]:=[a,b_1, b_2, \cdots b_n]
\end{equation}

Let $q_c,q_s\in k$ be two elements in a ring $k$. We define on
$k<A>=k[A^*]$ a new law $\ua$ by
\begin{eqnarray}\label{infiltr}
\hspace{-15mm} w\ua 1_{A^*}=1_{A^*}\ua w=w\cr \hspace{-15mm} a\ast u
\ua b\ast v=a\ast (u\ua b\ast v)+ q_c^{|a\ast u||b|} b\ast (a\ast
u\ua v)+ q_c^{|u||b|}q_s^{|a||b|} ab\ast (u\ua v)
\end{eqnarray}
where the weights ($|x|=n$ if $x\in A_n$) are extended additively to
lists by
$$
\Big|[a_1,a_2,\cdots ,a_k]\Big|=\sum_{i=1}^k |a_i|
$$
Then the new law $\ua$ is graded, associative with $1_{A^*}$ as unit.\\
\end{proposition}

\Proof It suffices to prove the identity $x\inf (y\inf z)=(x\inf
y)\inf z$ ; $x,y,z$ being lists (as the two members are trilinear).
It is obviously true when one of the factors is the empty list. We
show it when the three factors are non-empty (throughout the
computation, the law $*$ will have priority over other operators).

\begin{eqnarray}
(a\ast u\inf b\ast v)\inf c\ast w =\cr \big(a\ast (u\inf b\ast v)+
q^{|u||b|}t^{|a||b|} \2m{b}{a}(u\inf v)+q^{|a\ast u||b|} b(a\ast
u\inf v)\big)\inf c\ast w =\cr \Big[a\ast ((u\inf b\ast v)\inf c\ast
w)+q^{(|u|+|b\ast v|)|c|}t^{|a||c|}\2m{c}{a}((u\inf b\ast v)\inf
w)\cr +q^{(|a\ast u|+|b\ast v|)|c|}c\ast (a\ast (u\inf b\ast v)\inf
w) \Big] +\cr \Big[q^{|u||b|}t^{|a||b|} \2m{b}{a}(u\inf v\inf c\ast
w)
+q^{|u||b|+(|u|+|v|)|c|}t^{|a||b|}t^{(|a|+|b|)|c|}\3m{c}{b}{a}(u\inf
v\inf\ w)+\cr q^{|u||b|+(|a\ast u|+|b\ast v|)|c|}t^{|a||b|}c
((\2m{b}{a}(u\inf v))\inf w)\Big]+\cr \Big[q^{|a\ast u||b|}b((a\ast
u\inf v)\inf c\ast w)+q^{|a\ast u||b|+ (|a\ast
u|+|v|)|c|}t^{|b||c|}\2m{c}{b}(au\inf v\inf w)+\cr q^{|a\ast
u||b|+(|a\ast u|+|b\ast v|)|c|}c(b(a\ast u\inf v)\inf w)\Big]
\end{eqnarray}

\begin{eqnarray}
a\ast u\inf (b\ast v\inf c\ast w)=\cr a\ast u\inf \big(b\ast (v\inf
c\ast w)+q^{|v||c|}t^{|b||c|} \2m{c}{b}(v\inf w)+q^{|b\ast v||c|}
c(b\ast v\inf w)\big)=\cr \Big[a\ast (u\inf b\ast (v\inf c\ast w))+
q^{|u||b|}t^{|a||b|}\2m{b}{a}(u\inf v\inf c\ast w)+q^{|a\ast
u||b|}b(a\ast u\inf v\inf c\ast w)\Big]+\cr
\Big[q^{|v||c|}t^{|b||c|}a\ast (u\inf \2m{c}{b}(v\inf w))
+q^{|v||c|+|u|(|c|+|b|)}t^{|b||c|+|a|(|b|+|c|)}\3m{c}{b}{a}(u\inf
v\inf w)+\cr q^{|v||c|+|a\ast u|(|b|+|c|)}t^{|b||c|}\2m{c}{b}(a\ast
u\inf v\inf w)\Big]+\cr \Big[q^{|b\ast v||c|}a\ast (u\inf c(b\ast
v\inf w))+q^{(|u|+|b\ast v|)|c|}t^{|a||c|}\2m{c}{a}(u\inf b\ast
v\inf w)+\cr q^{(|a\ast u|+|b\ast v|)|c|}c\ast (a\ast u\inf b\ast
v\inf w)\Big]
\end{eqnarray}
in the second expression, one gathers the three terms which we find
first in the square brackets and we get

\begin{eqnarray}
a\ast (u\inf b\ast (v\inf cw))+q^{|v||c|}t^{|b||c|}a\ast (u\inf
\2m{c}{b}\ast (v\inf w))+\cr q^{|b\ast v||c|}a\ast (u\inf c\ast
(b\ast v\inf w))=a\ast (u\inf b\ast v\inf c\ast w)
\end{eqnarray}
in the first expression, one gathers the three terms which we find
last in the square brackets and we get

\begin{eqnarray}
q^{(|a\ast u|+|b\ast v|)|c|}c\ast (a\ast (u\inf b\ast v)\inf w) +\cr
q^{|u||b|+(|a\ast u|+|b\ast v|)|c|}t^{|a||b|} c \ast ((\2m{b}{a}\ast
(u\inf v))\inf w) +\cr q^{|a\ast u||b|+(|a\ast u|+|b\ast
v|)|c|}c\ast (b\ast (a\ast u\inf v)\inf w)=\cr
q^{(|au|+|bv|)|c|}c\ast (a\ast u \inf b\ast v \inf w)
\end{eqnarray}
and one finds the 7-term expression

\begin{eqnarray}
a\ast (u\inf b\ast v \inf c\ast w) + q^{|a\ast u|}b\ast (a\ast u\inf
v\inf c\ast w) +\cr q^{|a\ast u|+|b\ast v|}c\ast (a\ast u\inf b\ast
v\inf w)+
 q^{|u||b|}t^{|a||b|} \2m{b}{a}\ast (u\inf v\inf c\ast w)+\cr
 q^{(|u|+|b\ast v|)|c|}t^{|a||c|} \2m{c}{a}\ast (u\inf b\ast v\inf w)+\cr
 q^{|v||c|(|b|+|c|)|au|}t^{|b||c|} \2m{c}{b}\ast (a\ast u\inf v\inf w)\cr
+ q^{|v||c|+|u|(|c|+|b|)}t^{|b||c|+|a|(|b|+|c|)}\3m{c}{b}{a}\ast
(u\inf v\inf w)
\end{eqnarray}

\cqfd

The framework with diagrams will need another proposition on shifted
laws.

\subsection*{$\bullet$ Shifting}

We begin by the ``shifting lemma''.

\begin{lemma}
Let $\A$ be an associative algebra (whose law will be denoted by $\star$) and \\
$\A=\oplus_{n\in \N} \A_n$ a decomposition of $\A$ in direct sum.
Let $T\in\mathrm{End}(\A)$ be an endomorphim of the algebra $\A$. We
will denote by $T^n=T\circ T\circ \cdots \circ T$ the n-th
compositional power of $T$. We suppose that the shifted law
\begin{equation}\label{shifted}
    a\ \bar\star\ b=a\star T^\al(b)
\end{equation}
for $a\in \A_\al$ is graded for the decomposition $\A=\oplus_{n\in \N} \A_n$.\\
Then, if the law $\star$ is associative so is the law $\bar\star$.
\end{lemma}

\begin{rem} The hypothesis that the shifted law given by eq.\mref{shifted} be graded is automatically satisfied if $\A=\oplus_{n\in \N} \A_n$ is a graded algebra and if all the morphisms $T_n$ are of degree $0$.
\end{rem}

This lemma will be applied to the decomposition given by
$n=sup(Alph(w))$ (the highest index of variables appearing in $w$)
and the morphism given by $T(x_i)=x_{i+1}$.

What do these statements mean for us ?\\
Here the graded semigroup is $\mathfrak{MON}^+(X)$ and we do not forget
the coding arrow $\varphi_{lm}: \ldiag \ra (\mathfrak{MON}^+(X))^*$. The
image of $\varphi_{lm}$ is exactly the set of lists of monomials
$w=[m_1,m_2,\cdots ,m_k]$ such that the set of variables involved
$Alph(w)$ is of the form $x_1\cdots x_l$ (the labelling of the white
spots is without hole). By abuse of language we will say that a list
of monomials ``is in $\ldiag$'' in this case. It is not difficult to
see, from formulas (\ref{infiltr},\ref{shifted}) that if $w_i,\
i=1,2$ are in $\ldiag$ so are all the factors of $w_1\bar\ua w_2$,
this defines a new law on $K[\ldiag]$ and this algebra will be
called $\LDIAG(q_c,q_s)$. The properties of this algebra will be
made precise in the following proposition.

\begin{proposition} Let $\mathcal{C}_{ldiag}$ be the subspace of $(K<\mathfrak{MON}^+(\X)>,\bar\ua)$ generated by the codes of the diagrams (i.e. the lists $w\in \mathfrak{MON}^+(\X)$ such that $Alph(w)$ is without hole). Then\\
i) $(\mathcal{C}_{ldiag},\bar\ua)$ is a unital subalgebra of $(K<\mathfrak{MON}^+(\X)>,\bar\ua)$\\
ii) $(\mathcal{C}_{ldiag},\bar\ua)$ is a free algebra. More
precisely, for any diagram decomposed in irreducibles
$d=d_1.d_2\cdots d_k$ let
\begin{equation}
B(d):=\varphi_{lm}(d_1)\bar\ua\varphi_{lm}(d_2)\cdots
\bar\ua\varphi_{lm}(d_k)
\end{equation}
then \\
$\al$) $(B(d))_{d\in ldiag}$ is a basis of $\mathcal{C}_{ldiag}$\\
$\be$) $B(d_1.d_2)=B(d_1)\bar\ua B(d_2)$
\end{proposition}

As $k[\ldiag]$ is isomorphic to $\mathcal{C}_{ldiag}$ as a linear
space, we denote $\LDIAG (q_c,q_s)$ the new algebra structure of
$k[\ldiag]$ inherited from $\mathcal{C}_{ldiag}$. one has
\begin{equation}
    \LDIAG (0,0)\simeq \LDIAG;\ \LDIAG (1,1)\simeq \MQS
\end{equation}

\section{Coproducts}

We must now define a parametrized (say, by $q_t$) coproduct such that\\
$(\LDIAG(q_c,q_s),\bar\ua, 1_{\ldiag},\Delta_{q_t},\ep)$ is a graded bialgebra (as in the non-deformed Hopf algebra of \cite{GOF4}, the counit $\ep$ is just the ``constant term'' linear form).\\
We will take advantage of the freeness of $\LDIAG(q_c,q_s)$ through
the following lemma.

\begin{lemma}\label{coprodvar} Let $\Y$ be an alphabet, $k$ a ring and\\
$k<\Y>=k[\Y^*]$ be the free algebra constructed on $\Y$. For every mapping\\
$\Delta : A\ra k<\Y>\otimes k<\Y>$, we denote $\bd : k<\Y>\mapsto
k<\Y>\otimes k<\Y>$ its extension as a morphism of algebras
($k<\Y>\otimes k<\Y>$ being endowed with its non-twisted structure
of tensor product of algebras). Then, in order to be coassociative, it
is necessary and sufficient that

\begin{equation}
    (\bd\otimes I)\circ \Delta\textrm{ and } (I\otimes \bd)\circ \Delta
\end{equation}
 coincide on $\Y$.
\end{lemma}

The preceding lemma expresses the fact that, for a free algebra, the variety
 of the possible coproducts is a linear subspace. This will
be transparent in formula \mref{tline}.

\ss We now consider the structure constants of the coproduct of
$\MQS$ \cite{DHT} expressed with respect to the family of free
generators
$$
    \{MS_P\}_{P\in \mathcal{PM}^c}
$$
where $\mathcal{PM}^c$ is the set of connex packed matrices
(similarly, $\mathcal{PM}$ is the set of packed matrices).

\begin{equation}
\Delta_{\MQS}(MS_P)=\sum_{Q,R\in \mathcal{PM}}\al_P^{Q,R}\
MS_Q\otimes MS_R
\end{equation}

For the irreducible diagram $d$, we set

\begin{equation}
    \Delta_1(d)=\sum_{d_1,d_2\in irr(\ldiag)} \al_{\varphi_{lm}(d)}^{\varphi_{lm}(d_1),\varphi_{lm}(d_2)} d_1\otimes d_2
\end{equation}

and $\Delta_0(d)=\Delta_{WS}(d)$. Then proposition \mref{coprodvar}
proves that, for $q_t\in \{0,1\}$

\begin{equation}\label{tline}
    \Delta_t=\overline{(1-q_t)\Delta_0+q_t\Delta_1} 
\end{equation}
is a coproduct of graded bialgebra for
$(\LDIAG(q_c,q_s),\ua,1_{\ldiag})$.

\ss We sum up the results

\begin{proposition}
i) With the operations defined above, $q_c,q_s$ complex or formal and $q_t$ boolean ($q_t\in \{0,1\}$), 
$$
\LDIAG(q_c,q_s,q_t):=\left(\LDIAG(q_c,q_s),\bar\ua,1_{\ldiag},\Delta_{q_t},\ep\right)
$$
is a Hopf algebra.\\
ii) At parameters $(0,0,0)$, one has $\LDIAG(0,0,0)\simeq \LDIAG$\\
iii) At parameters $(1,1,1)$, one has $\LDIAG(1,1,1)\simeq \MQS$
\end{proposition}

\section{More on $\LDIAG(q_c,q_s,q_t)$ : structure, images and the link with Euler-Zagier sums}

It has been proved recently that $\LDIAG(q_c,q_s,q_t)$ is a {\em tridendriform} Hopf Algebra \cite{Fo3} and that $\LDIAG(1,q_s,q_t)$ is a homomorphic image of the algebra of planar decorated trees of Foissy \cite{Fo1,Fo2}. {\em Bidendriformity} of the algebra $\LDIAG(q_c,q_s)$ can also be established through a bi-word realization providing yet another (statistical) interpretation of the $(q_c,q_s)$ deformation \cite{FPSAC07}.\\
We will now make clear the relations between the $(q_c,q_s)$ deformation and  Euler-Zagier sums.\\
According the notation of \cite{Kr}, one has
\begin{equation}
    \zeta(s_1,\cdots, s_n;\si_1,\cdots,\si_n)=\sum_{0<i_1<\cdots <i_n}
    \frac{\si_1^{i_1}\cdots \si_n^{i_n}}{i_1^{s_1}\cdots i_n^{s_n}}
\end{equation}
with $\si_i\in\{-1,1\}$ and $s_1>1$ if $\si_1=1$. Here we are more
interested in the multiplication mechanism, so we extend the
notation to formal variables and use, for indices, the bi-word
notation. Hence
\begin{equation}
    \zeta_{FP}\left(\,\matrix{m_1 & \cdots & m_n \cr s_1 & \cdots & s_n}\,\right)=\sum_{0<i_1<\cdots <i_n}
    \frac{m_1^{i_1}\cdots m_n^{i_n}}{i_1^{s_1}\cdots i_n^{s_n}}.
\end{equation}
We remark that the indices are taken as words (i.e. lists) with
variables located in the semigroup $\mathfrak{MON}(Z)\times \N^+$ with
$Z=\{z_i\}_{i\geq 1}$. The set of these functions is closed under
multplication and will be called below $FP(Z), \textit{\ formal
polyzeta functions in the variables}\ Z$. Hence, the multiplication of these
sums fits in the hypotheses of Proposition \mref{2params} with
$q_c=q_s=1$ (quasi-shuffle in \cite{Ca3}). From this, we deduce an
arrow
\begin{equation}\label{poly_arrow}
LDIAG(1,1)\ra FP(Z).
\end{equation}
More precisely, if $d$ is a diagram with code $[m_1,m_2\cdots ,m_p]$
we make correspond
\begin{equation}\label{image_poly}
    \zeta_{FP}\left(\,\matrix{m_1 & \cdots & m_n \cr \deg{m_1} & \cdots & \deg{m_n}}\,\right)
\end{equation}
where $\deg{m_i}$ is the total degree of $m_i$. We will denote $\zeta_{D2FP}(d)$ this value \mref{image_poly}.\\
One has
\begin{equation}
\zeta_{D2FP}(d_1)\zeta_{D2FP}(d_2)=\zeta_{D2FP}(d_1\ua_{11} d_2)
\end{equation}
the law $\ua_{11}$ being unshifted and specialized to $(q_c,q_s)=(1,1)$.\\
When restricted to ``convergent'' diagrams (i.e. diagrams with
$\deg{m_1}\geq 2$ which form a subalgebra of $\LDIAG_u(q_c,q_s)$)
and specializing all the variables to $1$, we recover the ``usual''
Euler-Zagier sums by just counting the outgoing degrees of the black
spots and the arrow of \mref{poly_arrow} becomes
\begin{equation}\label{usual_EZ}
d\ra    \zeta(\deg{m_1},  \cdots  ,\deg{m_n})
\end{equation}
(usual Euler-Zagier sums). Denoting the last \mref{usual_EZ} value
$\zeta_{D2EZ}(d)$, one has
\begin{equation}
\zeta_{D2EZ}(d_1)\zeta_{D2EZ}(d_2)=\zeta_{D2EZ}(d_1\ua_{11} d_2)
\end{equation}

\section{Concluding remarks}

For a diagram $d$ with $r$ black spots, the code $[m_1,m_2,\cdots
,m_r]$ can be temporarily seen as a ``vector of coordinates'' for
the given diagram, but we prefer to stick to the structure of lists
as, firstly, the dimension of the vector varies with the diagram and
secondly, we have to concatenate the codes. The coordinate functions
of the diagram $d$ are therefore the family $(a_i)_{i>0}$ defined by
$a_i(d)=m_i$ for $i\leq r$ and $a_i(d)=0$ for $i>r$. From this
perspective the ``$q_t$'' of our three parameter deformation is a
quantization in the sense of Moyal's deformed products \cite{BFFLS}
on the algebra of coordinate functions (but without the first order
condition; see the introduction of \cite{CP}), by the
formula
\begin{equation}
    a_{i_1}*a_{i_1}\cdots *a_{i_k}(d)=\mu(a_{i_1}\otimes a_{i_1}\otimes \cdots \otimes a_{i_k}(\Delta_{q_t}^{[k]}(d)))
\end{equation}
 where $\mu$ is the ordinary multiplication of polynomials.\\
 The crossing parameter $q_c$ is also a quantization parameter as, for $q_s=0$, one has

\begin{equation}
    code(d_1*d_2)=code(d_1)\sqcup_{q_c}T(code(d_2))
\end{equation}
 where $T$ is a suitable translation of the variables and $\sqcup_{q_c}$ is the quantum shuffle \cite{Ro} for the braiding on $V=\C[x_i;\ i\geq 1]$ defined by

\begin{equation}
\hspace{-1.7cm} B(x_{i_1}^{\al_1}x_{i_2}^{\al_2}\cdots
x_{i_k}^{\al_k}\otimes y_{j_1}^{\be_1}y_{j_2}^{\be_2}\cdots
y_{j_l}^{\be_l})= q_c^{(\sum \al_i)(\sum \be_j)}\
y_{j_1}^{\be_1}y_{j_2}^{\be_2}\cdots y_{j_l}^{\be_l}\otimes
x_{i_1}^{\al_1}x_{i_2}^{\al_2}\cdots x_{i_k}^{\al_k}
\end{equation}

Let us add that $q_s$ and $q_c$ are of different nature as $q_s$
is the coefficient of a perturbation of the shuffle product (better
seen on the coproduct). This kind of perturbation occurs in various
domains as : computer science by means of the infiltration product
introduced by Ochsenschl\"ager \cite{Oc} (see also \cite{DL} and
\cite{DFLL}), algebra of the Euler-Zagier sums \cite{Ho} and
noncommutative symmetric functions \cite{DHT}. The mathematics of
this dual aspect is of geometrical nature and will be developed in
\cite{DKPTT}.

\end{document}

%% file: unlabelled2_1.tex
\ifx\JPicScale\undefined\def\JPicScale{1}\fi
\unitlength \JPicScale mm

\begin{picture}(122.5,87.5)(0,5)
\linethickness{0.75mm}
\multiput(69.34,49.61)(0.12,0.37){76}{\line(0,1){0.37}}
\linethickness{0.75mm}
\multiput(72.23,52.11)(0.12,0.37){67}{\line(0,1){0.37}}
\linethickness{0.75mm}
\multiput(82.76,78.82)(0.12,-0.57){48}{\line(0,-1){0.57}}

\put(60,80){\circle{5}}
\put(80.13,79.87){\circle{5}}
\put(100.13,79.87){\circle{5}}
\put(120,80){\circle{5}}

\put(70.2,49.88){\circle*{5}}
\put(90.3,49.56){\circle*{5}}
\put(110,50){\circle*{5}}

\put(56,86){$\{1\}$}
\put(70,86){$\{2,3,4\}$}
\put(88,86){$\{5,6,7,8,9\}$}
\put(115,86){$\{10,11\}$}

\put(60.2,41.88){$\{2,3,5\}$}
\put(77.3,41.88){$\{1,4,6,7,8\}$}
\put(104,41.88){$\{9,10,11\}$}

\linethickness{0.75mm}
\multiput(71.97,49.74)(0.12,0.14){212}{\line(0,1){0.14}}

\linethickness{0.75mm}
\multiput(102.76,79.47)(0.12,-0.48){57}{\line(0,-1){0.48}}

\linethickness{0.75mm}
\multiput(92.76,49.61)(0.12,0.36){81}{\line(0,1){0.36}}
\linethickness{0.75mm}
\multiput(90.26,48.29)(0.12,0.37){78}{\line(0,1){0.37}}
\linethickness{0.75mm}
\multiput(88.29,49.08)(0.12,0.36){81}{\line(0,1){0.36}}
\linethickness{0.75mm}
\multiput(111.97,50.66)(0.12,0.37){72}{\line(0,1){0.37}}
\linethickness{0.75mm}
\multiput(109.74,51.05)(0.12,0.37){73}{\line(0,1){0.37}}
\linethickness{0.75mm}
\multiput(61.71,78.03)(0.12,-0.13){216}{\line(0,-1){0.13}}
\end{picture}

\vspace{-2cm}

{\small{\bf Fig 1}\pointir \it Diagram from $P_1,\ P_2$ (set partitions of $[1\cdots 11]$).\\ 
$P_1=\left\{\{2,3,5\},\{1,4,6,7,8\},\{9,10,11\}\right\}$ and  $P_2=\left\{\{1\},\{2,3,4\},\{5,6,7,8,9\},\{10,11\}\right\}$ (respectively black spots for $P_1$ and white spots for $P_2$).\\ 
The incidence matrix corresponding to the diagram (as drawn) or these partitions is 
${\pmatrix{0 & 2 & 1 & 0\cr 1 & 1 & 3 & 0\cr 0 & 0 & 1 & 2}}$. But, due to the fact that the defining partitions are unordered, one can permute the spots (black and white, between themselves) and, so, the lines and columns of this matrix can be permuted. the diagram could be represented by the matrix ${\pmatrix{0 & 0 & 1 & 2\cr 0 &  2 & 1 & 0\cr 1 & 0 & 3 & 1}}$ as well.}


%% file: labelled2_1.tex
\ifx\JPicScale\undefined\def\JPicScale{1}\fi
\unitlength \JPicScale mm

\begin{picture}(122.5,47.5)(0,46)
\linethickness{0.75mm}
\multiput(69.34,49.61)(0.12,0.37){76}{\line(0,1){0.37}}
\linethickness{0.75mm}
\multiput(72.23,52.11)(0.12,0.37){67}{\line(0,1){0.37}}
\linethickness{0.75mm}
\multiput(82.76,78.82)(0.12,-0.57){48}{\line(0,-1){0.57}}

\put(60,80){\circle{5}}
\put(80.13,79.87){\circle{5}}
\put(100.13,79.87){\circle{5}}
\put(120,80){\circle{5}}

\put(70.2,49.88){\circle*{5}}
\put(90.3,49.56){\circle*{5}}
\put(110,50){\circle*{5}}

\put(59,86){$1$}
\put(79,86){$2$}
\put(99,86){$3$}
\put(119,86){$4$}

\put(69,41.88){$1$}
\put(89.3,41.88){$2$}
\put(109,41.88){$3$}

\linethickness{0.75mm}
\multiput(71.97,49.74)(0.12,0.14){212}{\line(0,1){0.14}}

\linethickness{0.75mm}
\multiput(102.76,79.47)(0.12,-0.48){57}{\line(0,-1){0.48}}

\linethickness{0.75mm}
\multiput(92.76,49.61)(0.12,0.36){81}{\line(0,1){0.36}}
\linethickness{0.75mm}
\multiput(90.26,48.29)(0.12,0.37){78}{\line(0,1){0.37}}
\linethickness{0.75mm}
\multiput(88.29,49.08)(0.12,0.36){81}{\line(0,1){0.36}}
\linethickness{0.75mm}
\multiput(111.97,50.66)(0.12,0.37){72}{\line(0,1){0.37}}
\linethickness{0.75mm}
\multiput(109.74,51.05)(0.12,0.37){73}{\line(0,1){0.37}}
\linethickness{0.75mm}
\multiput(61.71,78.03)(0.12,-0.13){216}{\line(0,-1){0.13}}
\end{picture}

\vspace{0.8cm}
\begin{center}
{\small{\bf Fig 2}\pointir \it Labelled diagram of format $3\times 4$ corresponding to the one of Fig 1.}
\end{center}

%% file: coding1.tex

\ifx\JPicScale\undefined\def\JPicScale{1}\fi
\def\JPicScale{0.7}
\psset{unit=\JPicScale mm}
\psset{linewidth=0.3,dotsep=1,hatchwidth=0.3,hatchsep=1.5,shadowsize=1,dimen=middle}
\psset{dotsize=0.7 2.5,dotscale=1 1,fillcolor=black}
\psset{arrowsize=1 2,arrowlength=1,arrowinset=0.25,tbarsize=0.7 5,bracketlength=0.15,rbracketlength=0.15}
\begin{pspicture}(10,0)(200.5,82.5)
\psline[linewidth=0.75](9.34,49.61)(18.42,77.89)
\psline[linewidth=0.75](20.26,76.84)(12.23,52.11)
\psline[linewidth=0.5,linestyle=dotted](22.76,78.82)(28.55,51.58)
\rput{0}(10.2,49.88){\psellipse[linewidth=0.75,fillstyle=solid](0,0)(2.5,2.5)}
\rput{0}(20.14,79.87){\psellipse[linewidth=0.75](0,0)(2.5,2.5)}
\rput{0}(30.3,49.56){\psellipse[linewidth=0.5,linestyle=dotted,fillstyle=solid](0,0)(2.5,2.5)}
\psline[linewidth=0.75](11.97,49.74)(37.37,79.21)
\rput{0}(50,50){\psellipse[linewidth=0.5,linestyle=dotted,fillstyle=solid](0,0)(2.5,2.5)}
\psline[linewidth=0.5,linestyle=dotted](42.76,79.47)(49.61,52.11)
\rput{0}(40.14,79.87){\psellipse[linewidth=0.75](0,0)(2.5,2.5)}
\rput{0}(0,80){\psellipse[linewidth=0.75](0,0)(2.5,2.5)}
\rput{0}(60,80){\psellipse[linewidth=0.75](0,0)(2.5,2.5)}
\psline[linewidth=0.5,linestyle=dotted](32.76,49.61)(42.5,79.08)
\psline[linewidth=0.5,linestyle=dotted](30.26,48.29)(39.61,76.84)
\psline[linewidth=0.5,linestyle=dotted](28.29,49.08)(38.03,78.55)
\psline[linewidth=0.5,linestyle=dotted](51.97,50.66)(60.66,77.37)
\psline[linewidth=0.5,linestyle=dotted](49.74,51.05)(58.55,77.89)
\psline[linewidth=0.5,linestyle=dotted](27.63,50.66)(1.71,78.03)
\psline[linewidth=0.5,linestyle=dotted](89.34,49.61)(98.42,77.89)
\psline[linewidth=0.5,linestyle=dotted](100.26,76.84)(92.23,52.11)
\psline[linewidth=0.75](102.76,78.82)(108.55,51.58)
\rput{0}(90.2,49.88){\psellipse[linewidth=0.5,linestyle=dotted,fillstyle=solid](0,0)(2.5,2.5)}
\rput{0}(100.13,79.87){\psellipse[linewidth=0.75](0,0)(2.5,2.5)}
\rput{0}(110.3,49.56){\psellipse[linewidth=0.75,fillstyle=solid](0,0)(2.5,2.5)}
\psline[linewidth=0.5,linestyle=dotted](91.97,49.74)(117.37,79.21)
\rput{0}(130,50){\psellipse[linewidth=0.5,linestyle=dotted,fillstyle=solid](0,0)(2.5,2.5)}
\psline[linewidth=0.5,linestyle=dotted](122.76,79.47)(129.61,52.11)
\rput{0}(120.13,79.87){\psellipse[linewidth=0.75](0,0)(2.5,2.5)}
\rput{0}(80.01,80){\psellipse[linewidth=0.75](0,0)(2.5,2.5)}
\rput{0}(140,80){\psellipse[linewidth=0.75](0,0)(2.5,2.5)}
\psline[linewidth=0.75](112.76,49.61)(122.5,79.08)
\psline[linewidth=0.75](110.26,48.29)(119.61,76.84)
\psline[linewidth=0.75](108.29,49.08)(118.03,78.55)
\psline[linewidth=0.5,linestyle=dotted](131.97,50.66)(140.66,77.37)
\psline[linewidth=0.5,linestyle=dotted](129.74,51.05)(138.55,77.89)
\psline[linewidth=0.75](107.63,50.66)(81.71,78.03)
\psline[linewidth=0.5,linestyle=dotted](169.34,49.61)(178.42,77.89)
\psline[linewidth=0.5,linestyle=dotted](180.26,76.84)(172.23,52.11)
\psline[linewidth=0.5,linestyle=dotted](182.76,78.82)(188.55,51.58)
\rput{0}(170.2,49.88){\psellipse[linewidth=0.5,linestyle=dotted,fillstyle=solid](0,0)(2.5,2.5)}
\rput{0}(180.14,79.87){\psellipse[linewidth=0.75](0,0)(2.5,2.5)}
\rput{0}(190.3,49.56){\psellipse[linewidth=0.5,linestyle=dotted,fillstyle=solid](0,0)(2.5,2.5)}
\psline[linewidth=0.5,linestyle=dotted](171.97,49.74)(197.37,79.21)
\rput{0}(210,50){\psellipse[linewidth=0.75,fillstyle=solid](0,0)(2.5,2.5)}
\psline[linewidth=0.75](202.76,79.47)(209.61,52.11)
\rput{0}(200.14,79.87){\psellipse[linewidth=0.75](0,0)(2.5,2.5)}
\rput{0}(160,80){\psellipse[linewidth=0.75](0,0)(2.5,2.5)}
\rput{0}(220,80){\psellipse[linewidth=0.75](0,0)(2.5,2.5)}
\psline[linewidth=0.5,linestyle=dotted](192.76,49.61)(202.5,79.08)
\psline[linewidth=0.5,linestyle=dotted](190.26,48.29)(199.61,76.84)
\psline[linewidth=0.5,linestyle=dotted](188.29,49.08)(198.03,78.55)
\psline[linewidth=0.75](211.97,50.66)(220.66,77.37)
\psline[linewidth=0.75](209.74,51.05)(218.55,77.89)
\psline[linewidth=0.5,linestyle=dotted](187.63,50.66)(161.71,78.03)
\end{pspicture}

\vspace{-2cm}
{\small{\bf Fig 3}\pointir \it Coding the diagram of fig 2 by a word of monomials. The code here is 
$[x_2^2x_3\ ,\ x_1x_2x_3^3\ ,\ x_3x_4^2]$}

%% file: GPE2.tex
\ifx\JPicScale\undefined\def\JPicScale{1}\fi
\psset{unit=\JPicScale mm}
\psset{linewidth=0.3,dotsep=1,hatchwidth=0.3,hatchsep=1.5,shadowsize=1,dimen=middle}
\psset{dotsize=0.7 2.5,dotscale=1 1,fillcolor=black}
\psset{arrowsize=1 2,arrowlength=1,arrowinset=0.25,tbarsize=0.7 5,bracketlength=0.15,rbracketlength=0.15}
\begin{pspicture}(0,0)(122.5,82.5)
\psline[linewidth=0.75](37.89,50.26)(29.74,77.37)
\psline[linewidth=0.75](32.11,78.16)(39.74,52.11)
\rput{0}(30.13,79.87){\psellipse[linewidth=0.75](0,0)(2.5,2.5)}
\rput{0}(40.3,49.56){\psellipse[linewidth=0.75,fillstyle=solid](0,0)(2.5,2.5)}
\rput{0}(50.13,79.87){\psellipse[linewidth=0.75](0,0)(2.5,2.5)}
\rput{0}(10,80){\psellipse[linewidth=0.75](0,0)(2.5,2.5)}
\rput{0}(70.01,80){\psellipse[linewidth=0.75](0,0)(2.5,2.5)}
\psline[linewidth=0.75](42.76,49.61)(52.5,79.08)
\psline[linewidth=0.75](40.26,48.29)(49.61,76.84)
\psline[linewidth=0.75](38.29,49.08)(48.03,78.55)
\psline[linewidth=0.75](43.42,50.79)(68.55,77.89)
\psline[linewidth=0.75](37.63,50.66)(11.71,78.03)
\psline[linewidth=0.75](87.89,50.26)(79.74,77.37)
\psline[linewidth=0.75](82.11,78.16)(89.74,52.11)
\rput{0}(80.13,79.87){\psellipse[linewidth=0.75](0,0)(2.5,2.5)}
\rput{0}(90.3,49.56){\psellipse[linewidth=0.75,fillstyle=solid](0,0)(2.5,2.5)}
\rput{0}(100.13,79.87){\psellipse[linewidth=0.75](0,0)(2.5,2.5)}
\rput{0}(120,80){\psellipse[linewidth=0.75](0,0)(2.5,2.5)}
\psline[linewidth=0.75](92.76,49.61)(102.5,79.08)
\psline[linewidth=0.75](90.26,48.29)(99.61,76.84)
\psline[linewidth=0.75](88.29,49.08)(98.03,78.55)
\psline[linewidth=0.75](93.42,50.79)(118.55,77.89)
\psline[linewidth=0.75](151.18,77.37)(159.21,51.85)
\rput{0}(150.14,79.87){\psellipse[linewidth=0.75](0,0)(2.5,2.5)}
\rput{0}(160.3,49.56){\psellipse[linewidth=0.75,fillstyle=solid](0,0)(2.5,2.5)}
\rput{0}(170.14,79.87){\psellipse[linewidth=0.75](0,0)(2.5,2.5)}
\rput{0}(130,80){\psellipse[linewidth=0.75](0,0)(2.5,2.5)}
\psline[linewidth=0.75](161.05,49.22)(170.53,77.37)
\psline[linewidth=0.75](159.6,48.82)(168.95,77.37)
\psline[linewidth=0.75](158.29,49.08)(168.03,78.55)
\psline[linewidth=0.75](162.89,50.53)(172.11,78.16)
\psline[linewidth=0.75](157.37,50.13)(131.45,77.5)
\psline[linewidth=0.75](158.55,51.84)(132.63,79.21)
\end{pspicture}

\vspace{-3cm}
{\small\it {\bf Fig 4}\pointir Graphic Primitive Elements of $\LDIAG$ have only one black spot and therefore are coded by the sequence of the ingoing degrees of their white spots (a composition). The first one here has code $[1,2,3,1]$. The picture shows an element of the monoid generated by Graphic Primitive Elements (a linear basis of $\LDIAG^{\textrm{\tiny GPE}}$) which is then coded by a list of compositions, here $\Big[[1,2,3,1],[2,3,1],[2,1,4]\Big]$.}

%% file: lbell.tex
\ifx\JPicScale\undefined\def\JPicScale{1}\fi
\psset{unit=\JPicScale mm}
\psset{linewidth=0.3,dotsep=1,hatchwidth=0.3,hatchsep=1.5,shadowsize=1,dimen=middle}
\psset{dotsize=0.7 2.5,dotscale=1 1,fillcolor=black}
\psset{arrowsize=1 2,arrowlength=1,arrowinset=0.25,tbarsize=0.7 5,bracketlength=0.15,rbracketlength=0.15}
\begin{pspicture}(-40,0)(122.5,92.5)
\psline[linewidth=0.75](21.32,76.84)(28.55,51.58)
\rput{0}(10.2,49.88){\psellipse[linewidth=0.75,fillstyle=solid](0,0)(2.5,2.5)}
\rput{0}(20.13,79.87){\psellipse[linewidth=0.75](0,0)(2.5,2.5)}
\rput{0}(30.3,49.56){\psellipse[linewidth=0.75,fillstyle=solid](0,0)(2.5,2.5)}
\psline[linewidth=0.75](10.53,50.53)(10.26,77.37)
\rput{0}(50,50){\psellipse[linewidth=0.75,fillstyle=solid](0,0)(2.5,2.5)}
\psline[linewidth=0.75](49.61,77.24)(49.61,52.11)
\rput{0}(40.13,79.87){\psellipse[linewidth=0.75](0,0)(2.5,2.5)}
\rput{0}(60.01,80){\psellipse[linewidth=0.75](0,0)(2.5,2.5)}
\psline[linewidth=0.75](31.32,50.79)(39.47,77.37)
\psline[linewidth=0.75](30,51.32)(30,77.11)
\psline[linewidth=0.75](50.53,50.53)(59.47,76.84)
\rput{0}(30,80){\psellipse[linewidth=0.75](0,0)(2.5,2.5)}
\rput{0}(10.13,79.87){\psellipse[linewidth=0.75](0,0)(2.5,2.5)}
\rput{0}(50.14,79.87){\psellipse[linewidth=0.75](0,0)(2.5,2.5)}
\end{pspicture}

\vspace{-3cm}
{\small\it {\bf Fig 5}\pointir An element of $\lbell$, concatenation $d_1d_3d_2$.}

%% file: cross1.tex
\ifx\JPicScale\undefined\def\JPicScale{1}\fi
\psset{unit=\JPicScale mm}
\psset{linewidth=0.3,dotsep=1,hatchwidth=0.3,hatchsep=1.5,shadowsize=1,dimen=middle}
\psset{dotsize=0.7 2.5,dotscale=1 1,fillcolor=black}
\psset{arrowsize=1 2,arrowlength=1,arrowinset=0.25,tbarsize=0.7 5,bracketlength=0.15,rbracketlength=0.15}
\begin{pspicture}(0,0)(233.56,82.5)
\psline[linewidth=0.75](8.29,50.79)(8.42,78.42)
\psline[linewidth=0.75](12.11,78.03)(11.84,51.18)
\psline[linewidth=0.75](12.89,78.95)(29.21,52.11)
\rput{0}(10.2,49.88){\psellipse[linewidth=0.75,fillstyle=solid](0,0)(2.5,2.5)}
\rput{0}(10.27,79.87){\psellipse[linewidth=0.75](0,0)(2.5,2.5)}
\rput{0}(30.3,49.56){\psellipse[linewidth=0.5,fillstyle=solid](0,0)(2.5,2.5)}
\rput{0}(50,50){\psellipse[linewidth=0.5,fillstyle=solid](0,0)(2.5,2.5)}
\rput{0}(49.74,80){\psellipse[linewidth=0.75](0,0)(2.5,2.5)}
\psline[linewidth=0.75](51.45,50.53)(51.32,77.76)
\psline[linewidth=0.75](48.03,51.71)(47.89,77.63)
\rput{0}(30,79.34){\psellipse[linewidth=0.75](0,0)(2.5,2.5)}
\psline[linewidth=0.75](28.68,77.5)(12.5,50.79)
\psline[linewidth=0.75](69.61,50.26)(69.74,77.89)
\psline[linewidth=0.75](73.42,77.5)(73.16,50.66)
\psline[linewidth=0.75](74.21,78.42)(90.53,51.58)
\rput{0}(71.52,49.35){\psellipse[linewidth=0.75,fillstyle=solid](0,0)(2.5,2.5)}
\rput{0}(71.58,79.34){\psellipse[linewidth=0.75](0,0)(2.5,2.5)}
\rput{0}(91.62,49.03){\psellipse[linewidth=0.5,fillstyle=solid](0,0)(2.5,2.5)}
\rput{0}(111.31,49.47){\psellipse[linewidth=0.5,fillstyle=solid](0,0)(2.5,2.5)}
\rput{0}(111.06,79.47){\psellipse[linewidth=0.75](0,0)(2.5,2.5)}
\psline[linewidth=0.75](112.76,50)(112.63,77.24)
\psline[linewidth=0.75](109.34,51.18)(109.21,77.11)
\rput{0}(91.58,79.08){\psellipse[linewidth=0.75](0,0)(2.5,2.5)}
\psline[linewidth=0.75](90,76.97)(73.82,50.26)
\psline[linewidth=0.75](129.61,50.26)(129.74,77.89)
\psline[linewidth=0.75](133.42,77.5)(133.16,50.66)
\psline[linewidth=0.75](134.21,78.42)(150.53,51.58)
\rput{0}(131.52,49.35){\psellipse[linewidth=0.75,fillstyle=solid](0,0)(2.5,2.5)}
\rput{0}(131.58,79.34){\psellipse[linewidth=0.75](0,0)(2.5,2.5)}
\rput{0}(151.62,49.03){\psellipse[linewidth=0.5,fillstyle=solid](0,0)(2.5,2.5)}
\rput{0}(171.06,79.47){\psellipse[linewidth=0.75](0,0)(2.5,2.5)}
\psline[linewidth=0.75](153.16,49.47)(172.63,77.24)
\psline[linewidth=0.75](151.58,51.58)(169.21,77.11)
\rput{0}(151.58,79.08){\psellipse[linewidth=0.75](0,0)(2.5,2.5)}
\psline[linewidth=0.75](150,76.97)(133.82,50.26)
\psline[linewidth=0.75](189.61,50.26)(189.74,77.89)
\psline[linewidth=0.75](193.42,77.5)(193.16,50.66)
\psline[linewidth=0.75](194.21,78.42)(210.53,51.58)
\rput{0}(191.52,49.35){\psellipse[linewidth=0.75,fillstyle=solid](0,0)(2.5,2.5)}
\rput{0}(191.58,79.34){\psellipse[linewidth=0.75](0,0)(2.5,2.5)}
\rput{0}(211.62,49.03){\psellipse[linewidth=0.5,fillstyle=solid](0,0)(2.5,2.5)}
\rput{0}(201.31,49.47){\psellipse[linewidth=0.5,fillstyle=solid](0,0)(2.5,2.5)}
\rput{0}(231.06,79.47){\psellipse[linewidth=0.75](0,0)(2.5,2.5)}
\psline[linewidth=0.75](203.16,49.74)(230.53,77.37)
\psline[linewidth=0.75](200,50)(228.16,78.42)
\rput{0}(211.58,79.08){\psellipse[linewidth=0.75](0,0)(2.5,2.5)}
\psline[linewidth=0.75](210,76.97)(193.82,50.26)
\psline[linewidth=0.75](78.29,8.95)(78.42,36.58)
\psline[linewidth=0.75](82.11,36.18)(81.84,9.34)
\psline[linewidth=0.75](82.89,37.11)(99.21,10.26)
\rput{0}(80.2,8.04){\psellipse[linewidth=0.75,fillstyle=solid](0,0)(2.5,2.5)}
\rput{0}(80.27,38.03){\psellipse[linewidth=0.75](0,0)(2.5,2.5)}
\rput{0}(100.3,7.72){\psellipse[linewidth=0.5,fillstyle=solid](0,0)(2.5,2.5)}
\rput{0}(119.74,38.16){\psellipse[linewidth=0.75](0,0)(2.5,2.5)}
\psline[linewidth=0.75](81.71,6.18)(119.21,36.05)
\psline[linewidth=0.75](82.76,9.34)(116.84,37.11)
\rput{0}(100.27,37.76){\psellipse[linewidth=0.75](0,0)(2.5,2.5)}
\psline[linewidth=0.75](98.68,35.66)(82.5,8.95)
\psline[linewidth=0.75](158.29,8.95)(158.42,36.58)
\psline[linewidth=0.75](162.11,36.18)(161.84,9.34)
\psline[linewidth=0.75](162.89,37.11)(179.21,10.26)
\rput{0}(160.2,8.04){\psellipse[linewidth=0.75,fillstyle=solid](0,0)(2.5,2.5)}
\rput{0}(160.27,38.03){\psellipse[linewidth=0.75](0,0)(2.5,2.5)}
\rput{0}(180.3,7.72){\psellipse[linewidth=0.5,fillstyle=solid](0,0)(2.5,2.5)}
\rput{0}(150,8.16){\psellipse[linewidth=0.5,fillstyle=solid](0,0)(2.5,2.5)}
\rput{0}(199.74,38.16){\psellipse[linewidth=0.75](0,0)(2.5,2.5)}
\psline[linewidth=0.75](152.11,7.37)(199.21,36.05)
\psline[linewidth=0.75](150.66,8.95)(196.84,37.11)
\rput{0}(180.27,37.76){\psellipse[linewidth=0.75](0,0)(2.5,2.5)}
\psline[linewidth=0.75](178.68,35.66)(162.5,8.95)
\end{pspicture}

\rput{0}(39.14,71.5){\Large\bf $*$}
\rput{0}(60.14,71.5){\Large\bf $=$}
\rput{0}(121.14,73.5){\Large\bf $+q_s^2$}
\rput{0}(178.14,73.5){\Large\bf $+\ q_c^2$}
\rput{0}(60.14,31){\Large\bf $+\ q_c^2q_s^6$}
\rput{0}(140.14,31){\Large\bf $+\ q_c^8$}

\bs
{\small\it {\bf Fig 5}\pointir Counting crossings and superposings produces an associative law.}

%% file: detail2.tex
\ifx\JPicScale\undefined\def\JPicScale{1}\fi
\psset{unit=\JPicScale mm}
\psset{linewidth=0.3,dotsep=1,hatchwidth=0.3,hatchsep=1.5,shadowsize=1,dimen=middle}
\psset{dotsize=0.7 2.5,dotscale=1 1,fillcolor=black}
\psset{arrowsize=1 2,arrowlength=1,arrowinset=0.25,tbarsize=0.7 5,bracketlength=0.15,rbracketlength=0.15}
\begin{pspicture}(-5,0)(101.35,110.71)
\psline[linewidth=0.75](28.55,50.79)(28.68,78.42)
\psline[linewidth=0.75](32.37,78.03)(32.11,51.18)
\psline[linewidth=0.75](33.16,78.95)(49.47,52.11)
\rput{0}(30.46,49.88){\psellipse[linewidth=0.75,fillstyle=solid](0,0)(2.5,2.5)}
\rput{0}(30.52,79.87){\psellipse[linewidth=0.75](0,0)(2.5,2.5)}
\rput{0}(50.57,49.56){\psellipse[linewidth=0.5,fillstyle=solid](0,0)(2.5,2.5)}
\rput{0}(70,80){\psellipse[linewidth=0.75](0,0)(2.5,2.5)}
\psline[linewidth=0.75](31.97,48.03)(69.47,77.89)
\psline[linewidth=0.75](33.03,51.18)(67.11,78.95)
\rput{0}(50.53,79.61){\psellipse[linewidth=0.75](0,0)(2.5,2.5)}
\psline[linewidth=0.75](48.95,77.5)(32.76,50.79)
\rput{0}(36.58,52.9){\psellipse[linewidth=0.65](0,0)(14.76,14.76)}
\psline[linewidth=0.75](122.96,54.98)(122.63,81.71)
\psline[linewidth=0.75](130,84.87)(129.97,55.73)
\psline[linewidth=0.75](146.45,85.53)(164.25,57.5)
\rput{0}(126.74,53.24){\psellipse[linewidth=0.75,fillstyle=solid](0,0)(4.86,4.86)}
\rput{0}(166.42,52.63){\psellipse[linewidth=0.5,fillstyle=solid](0,0)(4.87,4.87)}
\psline[linewidth=0.75](119.47,58.68)(157.76,79.87)
\psline[linewidth=0.75](117.11,61.84)(153.82,82.37)
\psline[linewidth=0.75](149.74,84.47)(131.27,54.98)
\rput{0}(138.45,58.24){\psellipse[linewidth=0.65](0,0)(28.54,28.54)}
\rput{0}(115,58.68){\psellipse[linewidth=0.75,fillstyle=solid](0,0)(4.86,4.86)}
\pspolygon[](112.76,73.16)(112.76,73.16)(112.76,73.16)(112.76,73.16)
\pspolygon[fillstyle=solid](131.28,65.46)(128.78,65.46)(128.98,62.7)(131.48,62.7)
\pspolygon[fillstyle=solid](146.81,79.28)(144.31,79.28)(144.51,76.51)(147.01,76.51)
\rput{0}(152.11,76.57){\psellipse[](0,0)(1.66,1.66)}
\rput{0}(150,80.12){\psellipse[](0,0)(1.66,1.66)}
\pspolygon[fillstyle=solid](126.01,55.2)(116.98,59.5)(115.75,56.89)(124.78,52.59)
\pspolygon[fillstyle=solid](142.86,71.91)(140.36,71.91)(140.56,69.14)(143.06,69.14)
\pspolygon[fillstyle=solid](124.18,61.65)(121.68,61.65)(121.88,58.88)(124.38,58.88)
\pspolygon[fillstyle=solid](124.18,66.65)(121.68,66.65)(121.88,63.88)(124.38,63.88)
\pspolygon[fillstyle=solid](129.18,56.65)(126.68,56.65)(126.88,53.88)(129.38,53.88)
\pspolygon[fillstyle=solid](131.28,70.33)(128.78,70.33)(128.98,67.56)(131.48,67.56)
\end{pspicture}

{\small\it {\bf Fig 6}\pointir Detail of the fourth monomial (with coefficient $q_c^2q_s^6$), crossings (circles) and superposings (black squares) are counted the same way but with a different variable.}